\documentclass[apjl]{emulateapj}

\usepackage{hyperref}
\usepackage{amsmath}
\usepackage{amssymb}
\usepackage{graphicx}
\usepackage{color}
\usepackage{natbib}

\newcommand{\MSun}{M_\odot}
\newcommand{\RSun}{R_\odot}
\newcommand{\Rstar}{R_*}
\newcommand{\Mstar}{m_*}
\newcommand{\MBH}{M_\mathrm{BH}}

\def\apj{Astrophysical Journal}                
\def\apjl{Astrophysical Journal}             

\def\mnras{MNRAS}            
\def\nat{Nature}              
\def\aap{A\&A}                
\def\na{New Astronomy}

\def\be{\begin{equation}}
\def\ee{\end{equation}}
\def\ba{\begin{eqnarray}}
\def\ea{\end{eqnarray}}

\begin{document}

\title{Double tidal disruptions in galactic nuclei}
\shorttitle{Double tidal disruptions}

\author{Ilya Mandel\altaffilmark{1,2} and Yuri Levin\altaffilmark{2}}
\affil{$^1$School of Physics and Astronomy, University of Birmingham, Birmingham, B15 2TT, United Kingdom\\
$^2$Monash Center for Astrophysics and School of Physics, Monash University, Clayton, VIC 3800, Australia}
\email{imandel@star.sr.bham.ac.uk}
\email{yuri.levin@monash.edu}

\begin{abstract}
A star on a nearly radial trajectory approaching a massive black hole (MBH) gets tidally disrupted if it comes sufficiently close to the MBH. Here we explore what happens to binary stars whose centers of mass approach the MBH on nearly radial orbits. The interaction with the MBH often leads to both stars being disrupted in sequence. We argue that such events could produce light curves that are substantially different from those of the single disruptions, with possible features such as two local maxima.   Tidal forces from the MBH can also lead the binary components to collide; these merger products can form highly magnetized stars, whose subsequent tidal disruption may enable prompt jet formation.
\end{abstract}



\section{Introduction}

Tidal disruption of stars by massive black holes (MBHs) in galactic nuclei has been studied by theorists for over 40 years, in the context of supplying gas for accretion onto SMBHs \citep[e.g.,][]{Hills:1975,Rees:1988,Phinney:1989}. During the past sixteen years, several tidal disruption event (TDE) candidates have been identified in the X-ray \citep[e.g.,][]{KomossaGreiner:1999,Komossa:2004,Levan:2011,Burrows:2011}, ultra-violet \citep[e.g.,][]{Gezari:2009,Bloom:2011}, optical \citep[e.g.,][]{vanVelzenFarrar:2011,Gezari:2012,Chornock:2014,Holoien:2014,Arcavi:2014} and radio \citep[e.g.,][]{Zauderer:2011} bands.  Observational progress has stimulated ongoing theoretical efforts to understand the details of tidal disruption physics and to model the lightcurves of TDEs \citep[e.g.,][]{Lodato:2009,StrubbeQuataert:2009,GuillochonRamirezRuiz:2013,GuillochonRamirezRuiz:2015,ShenMatzner:2014,Shiokawa:2015,Bonnerot:2015}.  The advent of deeper all-sky surveys such as LSST, as well as forthcoming transient surveys in the radio and ultra-violet bands, are anticipated to considerably increase the number of TDE observations.

The majority of the stars in the field, however, are in binaries, and some of the stars on a collisional course with the MBH will be members of a binary.  It is therefore interesting to consider an encounter between an MBH and a stellar binary, whose center of mass is approaching the MBH on a nearly parabolic orbit with the pericenter distance on the order of the tidal disruption radius of a single star.  In this paper we show that such encounters often lead to the tidal disruption of both stars in sequence. We find that when the encounter between a stellar binary and an MBH leads to a TDE, both members of the binary are disrupted in sequence more than 75\% of the time.
We estimate the frequency of such double TDEs, and argue that many of them will produce light curves that are different from those of single TDEs.   We discuss the implications of this scenario for recent observations of double flares in SDSS J015957.64+003310.5 \citep{Merloni:2015} and IC 3599 (Grupe et al 2015).

Collisions between binary components caused by tidal forces from the MBH are also common \citep{GinsburgLoeb:2007,Antonini:2010}.  They populate the phase space of low angular momentum stars with merger products many of which will eventually be tidally disrupted.  These merger products will have much larger magnetic fields than single stars \citep{Wickramasinghe:2014,Zhu:2015}.   The high magnetic flux is accreted onto the MBH during a TDE and can aid in generating prompt jets \citep{Tchekhovskoy:2014}, such as observed in Swift J164449.3+573451.

The structure of the paper is as follows.  In section \ref{sec:population} we develop a population model for the binaries approaching the MBH with small impact parameters, analytically justify why there should be a significant fraction of double TDEs, and describe the setup of numerical experiments.  In section \ref{sec:results} we present numerical experiments in which we follow the Newtonian dynamics of stars in binaries approaching the black hole, and flag tidal disruptions of individual stars and stellar collisions. In section \ref{sec:conclusions} we discuss the observational implications of our findings.

\section{Model of the binary population}\label{sec:population}

We are interested in binaries whose nearly radial approach to an MBH leads to a possible disruption of one or both binary companions. These binaries become tidally separated before they reach the point of closest approach \citep{Hills:1988,Sari:2010}.  

\begin{figure}
  \includegraphics[width=\columnwidth]{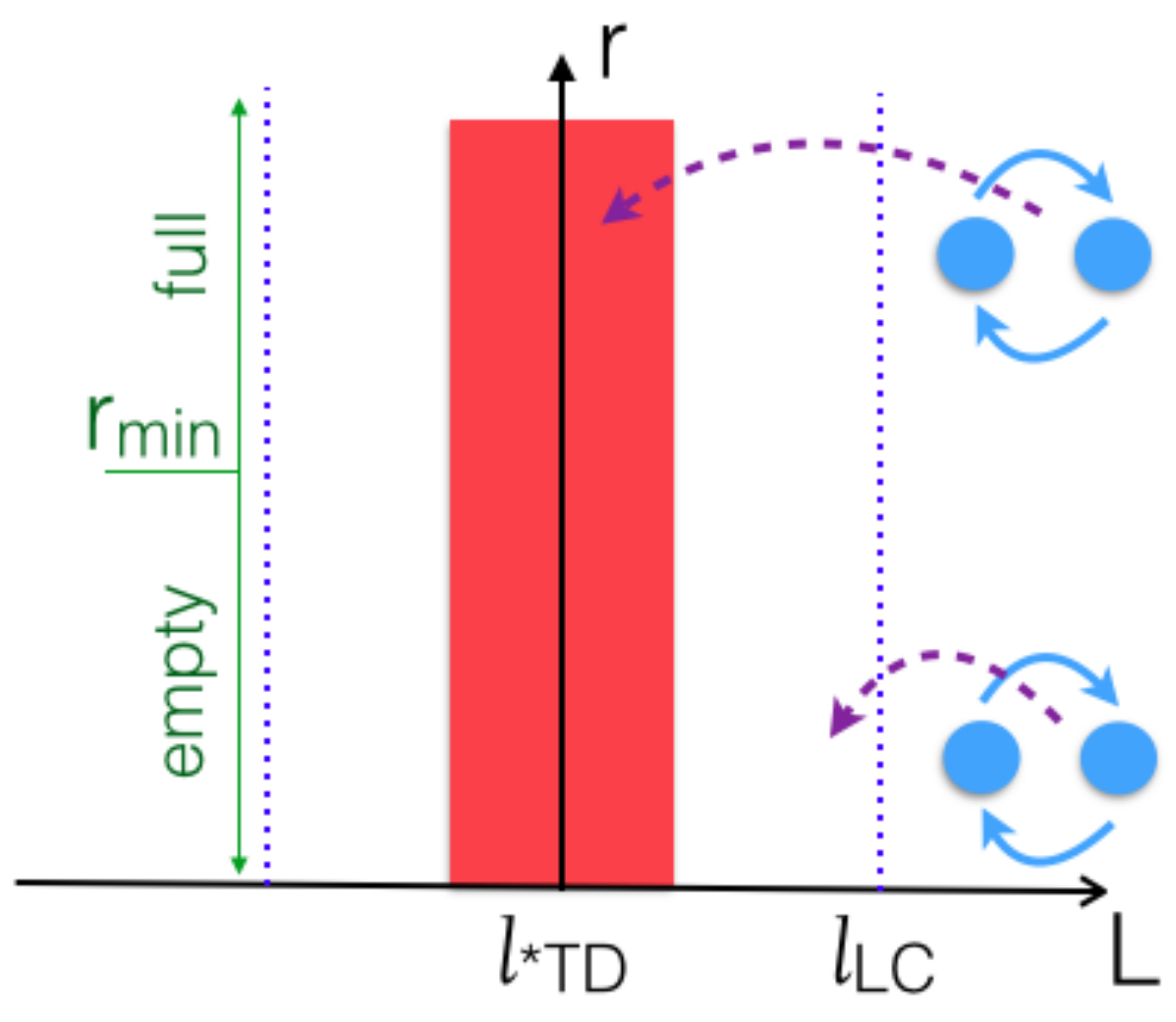}
  \caption{\label{fig:LC} An illustration of loss-cone dynamics.  The red rectangular region shows the angular momenta of individual stars which will be tidally disrupted, provided nothing changes their orbit on a dynamical timescale.  The blue dashed lines indicate the loss cone for tidal separation of stellar binaries (not to scale).  At small radial separations, $r \lesssim r_{min}$, the binary loss cone is empty, so binaries like the one on the bottom are tidally disrupted before entering the regime where individual stars can be disrupted; therefore, they cannot be the source of double TDEs.  At larger radii $r \gtrsim r_{min}$ (top binary), the loss cone is full and direct penetration into the stellar tidal disruption region is possible.  Double TDEs originate from the full binary loss cone region.} 
\end{figure}

A star of mass $\Mstar$ and radius $\Rstar$ will be tidally disrupted by an MBH of mass $\MBH$ if its orbit brings it within the tidal disruption radius of the MBH, 
\be
\label{rtidal}
R_\mathrm{*TD} = \alpha\Rstar (\MBH/\Mstar)^{1/3};
\ee
here $\alpha\sim 1$ depends on the structure of the star \citep{Phinney:1989, GuillochonRamirezRuiz:2013}. From here on we shall assume $\alpha=1$; however, some surviving stars will, in fact, lose a fraction of their mass \citep{GuillochonRamirezRuiz:2013}.  A binary on a nearly parabolic orbit around the MBH whose center of mass has a periapsis $r_p \lesssim R_\mathrm{*TD}$  is likely to first be tidally separated, and then to have both stars undergo sequential tidal disruptions.  This happens because the component stars have specific angular momenta that are close to that of the binary's center of mass, and thus undergo a similar periapsis approach.   The specific center of mass angular momentum of the binary on an orbit around the MBH with $r_p = R_\mathrm{*TD}$ is 
\be
l_\mathrm{*TD} \sim \sqrt{G \MBH \Rstar \left(\frac{\MBH}{\Mstar}\right)^{1/3}}.
\ee
The fractional difference in the angular momenta of the companions when the binary with semimajor axis $a$ is tidally separated is of order 
\be \label{deltal}
\frac{\delta l}{l_\mathrm{*TD}} \sim \frac{\sqrt{\frac{G \Mstar}{a}}\, a  \left(\frac{\MBH}{\Mstar}\right)^{1/3}} {l_\mathrm{*TD}}
\sim \sqrt{\frac{a}{\Rstar}} \left(\frac{\Mstar}{\MBH}\right)^{1/3},
\ee
i.e., less than one for $\Mstar = \MSun$, $\Rstar = \RSun$, $\MBH=10^6 \MSun$ and $a \lesssim 50$ AU. Therefore, the companion periapses will also typically be similar.

The phase space where angular momenta are sufficiently low to allow for stellar tidal disruptions is marked by a solid red rectangle in Fig.~\ref{fig:LC}. It is non-trivial for a binary to be dynamically inserted into this region. This is because binaries will be tidally separated by the MBH whenever the binary's angular momentum around the MBH is less than
\be
l_\mathrm{LC} \sim  \sqrt{G \MBH a \left(\frac{\MBH}{\Mstar}\right)^{1/3}},
\ee
a region bounded by the dotted blue lines in Fig.~\ref{fig:LC}.  Therefore, to produce a double TDE, a binary must enter the ``red'' region by jumping in from outside the blue lines!

Consider this dynamics in more detail.  Binaries wander stochastically in angular momentum space by gravitational scattering on other stars; this process can be accelerated by gravitational interaction with massive perturbers outside of the MBH's sphere of influence (SOI) \citep{Perets:2007} and by secular gravitational torques in the near-Keplerian potential deep inside the SOI \citep{RauchTremaine:1996}.   
If the typical change in angular momentum after one orbit is $\Delta l \ll l_\mathrm{LC}$, the binary can never get onto a small angular momentum orbit intact.  In the illustration of  Fig.~\ref{fig:LC}, the lower binary will be tidally separated immediately upon crossing the dotted blue line and will never reach the red rectangle; this regime is known as the empty loss cone \citep{LightmanShapiro:1977}.
Meanwhile, if $\Delta l \gtrsim l_\mathrm{LC}$, the binary can take large steps in angular momentum space, and can immediately jump anywhere within the loss cone $l <  l_\mathrm{LC}$; this is more probable for binaries coming from far away from the MBH, as illustrated by the top binary of Fig.~\ref{fig:LC}.  In the full loss cone regime, any angular momentum within the loss cone is nearly equally likely, and the probability that the binary ends up with a two-dimensional angular momentum $<|l_\mathrm{*TD}|$ is $\sim (l_\mathrm{*TD} / l_\mathrm{LC})^2$.  

The dividing line between the full and empty loss cone corresponds to $\Delta l \sim l_\mathrm{LC}$.  If the relaxation is governed by two-body interaction with other stars, the timescale for the angular momentum to change by the specific angular momentum of a circular orbit, $l_\mathrm{circ} = \sqrt{G \MBH r}$, is 
\be 
\tau_r = \frac{0.065 v^3}{G^2 \Mstar^2 n \log \Lambda}\, ,
\ee 
where $n(r)$ the local stellar number density and $v \sim \sqrt{G \MBH / r}$ the typical velocity at distance $r$ from the MBH \citep{SpitzerHart:1971}.  Assuming the stellar number density is a power law $n(r) = n_0 (r/r_0)^\gamma$, the loss cone is full for $r > r_{min}$ with
\be
r_{min} \sim \left[ \left(\frac{\MBH}{\Mstar}\right)^{7/3} \frac{0.065 r_0^\gamma}{2\pi n_0 \log \Lambda} a\right]^{1/(4+\gamma)}.
\ee 
At $r>r_{min}$, the ionization of binaries through interactions with other stars is unlikely \citep{Hopman:2009}.

The rate at which binaries enter the region where individual tidal disruptions are possible is roughly
\be
\label{d3N}
\frac{d^3 N}{dr da dt} (a,r) \sim   
\begin{cases}
 \frac{4\pi r^2 n(r) \pi (a)}{P} \left(\frac{l_\mathrm{*TD}}{l_\mathrm{circ}}\right)^2, & \text{if $r\gtrsim r_{min}$};\\
    0, & \text{otherwise},
  \end{cases}
\ee
where we take the density function of binary semi-major axes to be $\pi (a)= [\ln(a_{max}/a_{min})]^{-1} (1/a)$ \citep{Opik:1924}, and 
\be
P(r) = 2 \pi \sqrt{\frac{r^3}{G\MBH}}
\ee
is the period of the center of mass of the binary around the MBH.

The loss cone is ``half-full, half-empty'': roughly comparable numbers of binaries should enter it in the full and empty regimes \citep[][see \citep{CohnKulsrud:1978,MagorrianTremaine:1999} for rigorous analyses of loss-cone dynamics in galactic nuclei]{LightmanShapiro:1977}. 
Therefore, the rate of mergers per unit semimajor axis per unit time, i.e., the integral of Eq.~(\ref{d3N}) over $r$, can be roughly approximated by evaluating the integrand at $r=r_{min}$ and multiplying by $r_{min}$.  This yields the overall distribution of semimajor axes among binaries whose components can undergo individual TDEs immediately following the tidal separation of the binary:
\be \label{d2N}
\frac{d^2 N}{da dt} \propto a^{-7/(8+2\gamma)} = a^{-14/9}\, ,
\ee
where we assumed a Bahcall-Wolf cusp, $\gamma = -7/4$. Shallower cusps, such as the one in the Galactic center \citep{Bartko:2010}, would increase the fraction of double TDEs among all TDEs.
This estimate is simplistic, and does not take into account the detailed stellar distribution outside the SOI. However, a more sophisticated analysis performed by \citet{MacLeod:2012} in order to study the rate of tidal disruptions of giant stars is in agreement with our simple scaling. For a broad class of models, they found that the probability per unit time for a giant to be disrupted scales as $p(\Rstar) \propto \Rstar ^{0.35\pm0.15}$. Applying our calculation for the tidal separation of binaries, Eq.~(\ref{d3N}), to tidal disruptions, we would predict $p(\Rstar) \propto \Rstar^{4/9}$, in good agreement with \citep{MacLeod:2012}.  

Further integrating Eq.~(\ref{d3N}) over all $a$, we can obtain the total rate of binaries entering the  stellar disruption loss cone and compare this to the rate of individual TDEs of single stars, which is estimated by integrating the same equation with $a$ replaced by $\Rstar$.  Assuming the same density $n(r)$ of binary and single stars, and using the distribution choices discussed below, as many as  $(9/5) [\ln(a_{max}/a_{min})]^{-1} (a_{min}/\Rstar)^{-5/9} \sim 10\%$ of all TDEs could be contributed by double disruptions.

We now describe the details of our simulated population.  We fix the MBH mass to $\MBH = 10^6 \MSun$ in all cases.  The primary component mass is drawn from the Kroupa initial mass function \citep{Kroupa:2001} and the secondary is drawn from a distribution of the mass ratio $q=m_2/m_1$ given by $p(q) \propto q^{-3/4}$ for $q \in [0.2,1]$, broadly consistent with the observed field distributions of both low-mass \citep{DuquennoyMayor:1991} and high-mass \citep{Sana:2012,Sana:2013} stars.  We cut off the mass distribution for $q<0.2$ and $m_{1,2} < 0.1\, \MSun$, where it is poorly observationally constrained. 
  
As discussed above, binaries of interest are distributed according to $p(a) \propto a^{-14/9}$, so we simulate $p(P_b) \propto P_b^{-37/27}$ for binary periods $P_b$ between 0.1 day and 1000 years.  Approximately following \citet{DuquennoyMayor:1991}, we assume that binaries with periods less than 10 days are tidally circularized, binaries with periods above 1000 days have eccentricities drawn from a thermal distribution $p(e) = 2e$, and binaries with periods between 10 and 1000 days have their eccentricity drawn from a Gaussian distribution with mean 0.3 and standard deviation 0.15, cut off at 0 and 1.  We impose the additional constraint that the initial binary periapsis must be at least twice the sum of their radii, where the stellar radii are $\Rstar =(\Mstar/\MSun)^k \RSun$, with $k=0.8$ for $\Mstar<\MSun$ and $k=0.6$ for $\Mstar>\MSun$ \citep{KippenhahnWeigert:1994}.

We start each binary on a parabolic center-of-mass trajectory around the MBH at a distance ten times the binary tidal separation radius, with MBH periapses drawn uniformly from $\in [0,300]\ \RSun$ (Fig.~\ref{fig:rpa} demonstrates that no TDEs happen for larger periapses, in line with expectations, Eq.~(\ref{rtidal})). The three-body evolution is carried out with the \texttt{REBOUND} integrator \citep{ReinLiu:2012,ReinSpiegel:2015}.  The evolution proceeds until either (i) the components collide (the distance between the stars becomes smaller than the sum of their radii; we label all such events as `mergers' below although individual stars may survive in some cases \citep{Antonini:2011}); or (ii) one of the components is tidally disrupted (comes within a distance $R_\mathrm{*TD}$ of the MBH), after which the future trajectory of the remaining companion can be computed analytically as a 2-body problem; or (iii) the binary companions fly past the MBH without being disrupted or merging.  

\section{Results of numerical experiments}\label{sec:results}

We numerically simulated the dynamics of 1000 binaries as described above.  Among these, 18\% resulted in sequential tidal disruptions of both binary components and 5\% resulted in only one component (typically the more massive one) being disrupted.  Additionally, binary components merged in 6\% of simulations before either was disrupted; more mergers will arise from larger initial periapsis radii than considered here. The remaining 71\% of simulations produced no disruptions or mergers.

\begin{figure}
  \includegraphics[width=\columnwidth]{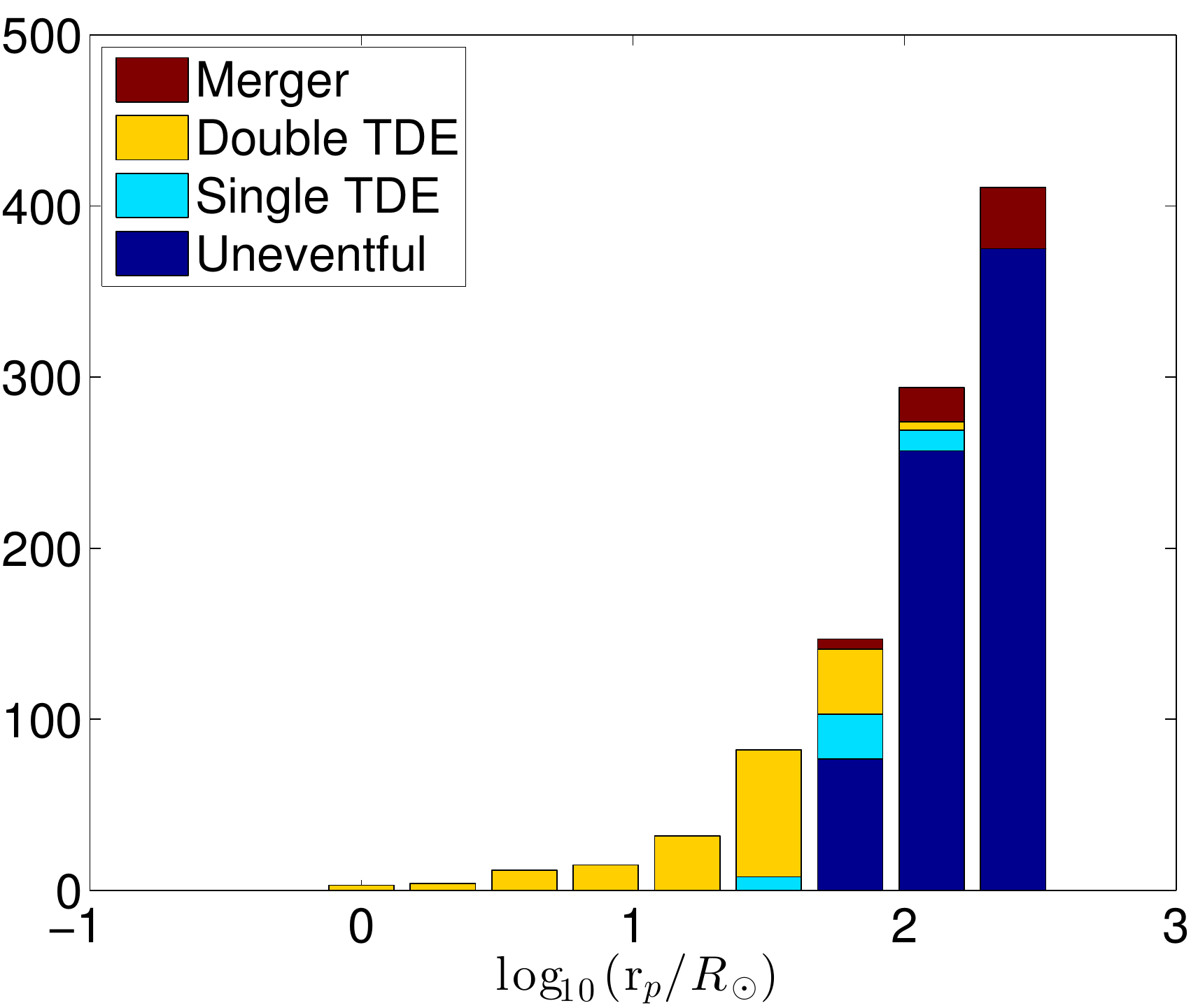}
  \includegraphics[width=\columnwidth]{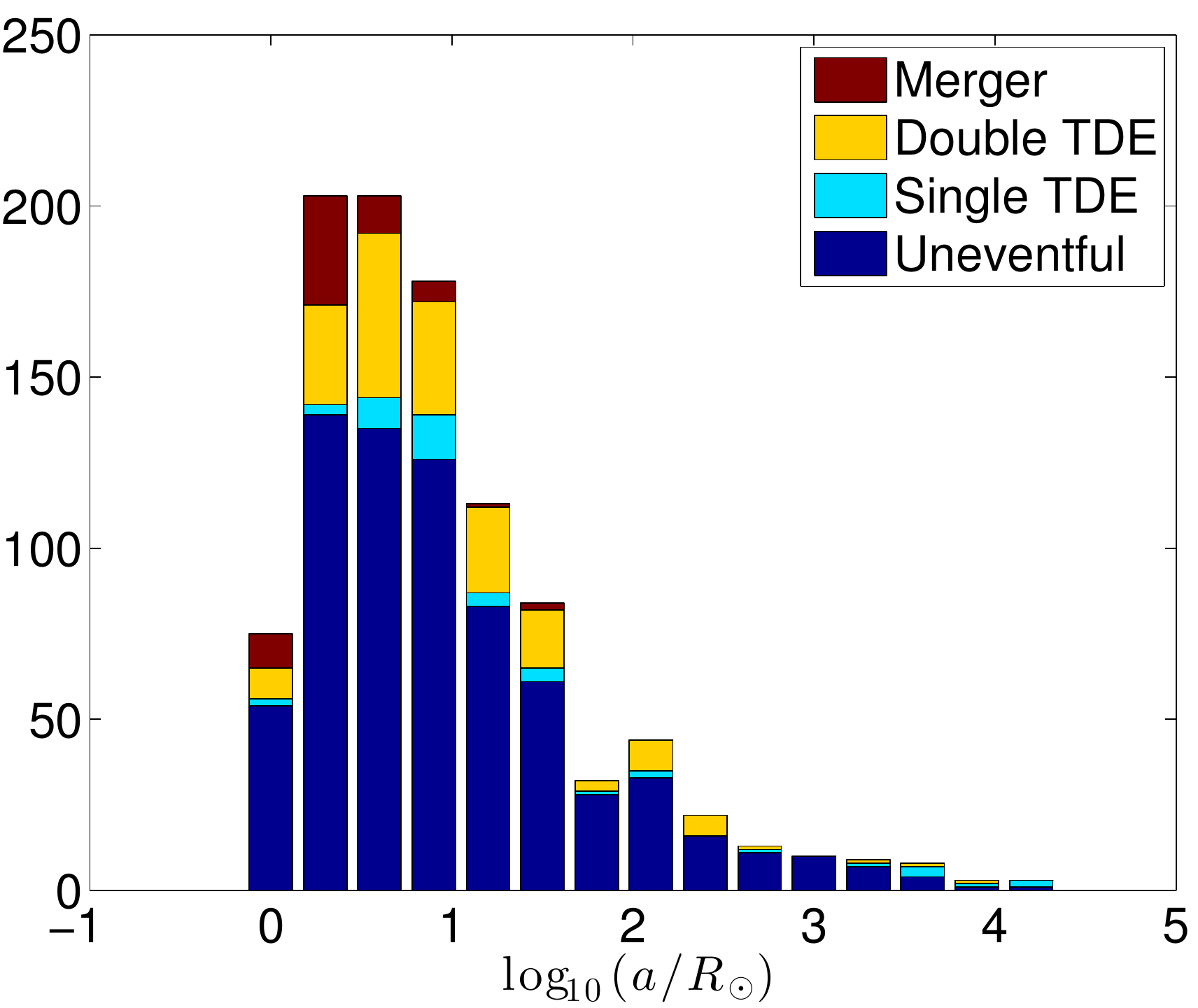}
  \caption{\label{fig:rpa} Distribution of mergers, double disruptions, single disruptions, and uneventful fly-bys (from top down) among the 1000 numerical simulations drawn from the distribution described in Section \ref{sec:population}, as a function of the center-of-mass periapsis $r_p$ around the MBH (top) and binary semi-major axis $a$ (bottom).} 
\end{figure}

Figure \ref{fig:rpa} shows the relative frequency of the possible outcomes as a function of the center-of-mass periapsis around the MBH and the initial binary semi-major axis.   Both components are likely to be disrupted if the periapsis is sufficiently small; see discussion around Eq.~(\ref{deltal}).  Single TDEs are relatively more common in comparison to double TDEs for large periapses, where they are correlated with significant companion mass differences which lead to different tidal disruption radii.   Mergers are relatively common for small binary separations, as tidal perturbations to the orbits are more likely to lead to collisions between companions; mergers preferentially happen for larger periapses, where disruptions are avoided. 

\begin{figure}
  \includegraphics[width=\columnwidth]{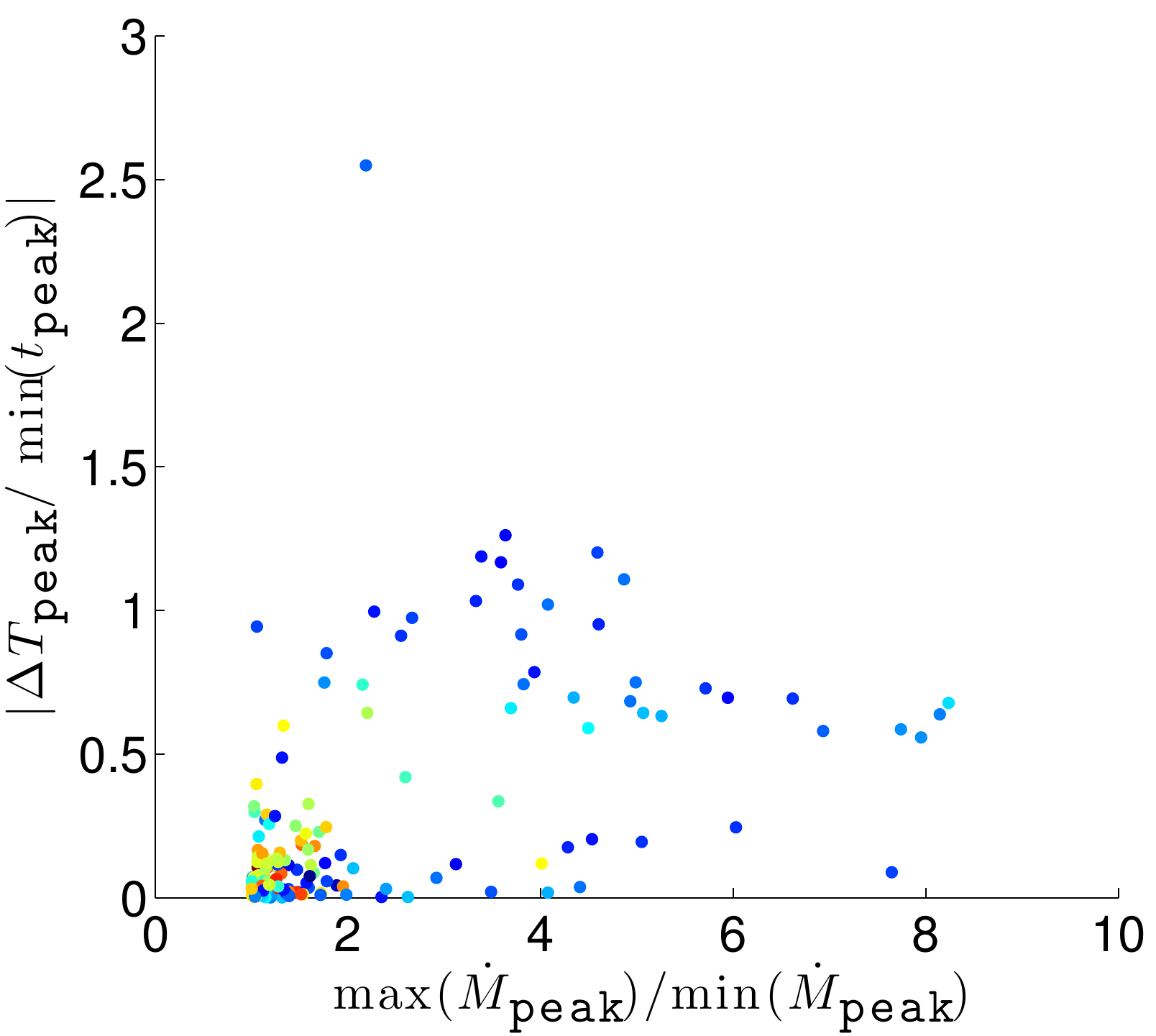}
  \caption{\label{fig:Deltat} Double TDEs: the ratio of the difference between peak times and the shorter rise time, vs. the ratio of the peak mass accretion rates. The color (red is 80 days, blue is 20 days) represents the shorter rise time.}
\end{figure}

When both components are tidally disrupted, we can ask whether individual events are resolvable.  The difference in peak luminosity times, $\Delta T_\mathrm{peak}$, is the difference in the rise times $t_\mathrm{peak}$ to the maximum value of the accretion rate $\dot{M}_\mathrm{peak}$, estimated using the fitting formulae of \citet{GuillochonRamirezRuiz:2013}, corrected for the the difference in the times when the binary components reach their periapses.  An additional correction of order $\Rstar/a$ to peak times and mass accretion rates, due to the energy shift of individual stellar orbits away from the zero-energy parabolic orbit, is not included.  Figure \ref{fig:Deltat} shows a scatter plot of the ratio of $\Delta T_\mathrm{peak}$ to the smaller of the two rise times, against the ratio of peak accretion rates for the two disrupted companions.  We see that peak accretion rates are typically comparable (the ratios are within a factor of two in $>70\%$ of cases), and never exceed $10:1$.  For 20\% of double TDEs, peak accretion times are separated by at least half the smaller rise time; even larger time delays $\Delta T_\mathrm{peak} > 2$ months with individual rise times of $t_\mathrm{peak} < 1$ month are observed.  The eccentricity vectors of the two component orbits, which point to the periapses, are nearly perfectly aligned in all cases, but orbital plane misalignments are occasionally possible, though rare (only 1 of 183 simulated double TDEs has orbital planes misaligned by more than 90 degrees) and generally correlated with larger $\Delta T$.

\begin{figure}
  \vspace{0.05in}
  \includegraphics[width=\columnwidth]{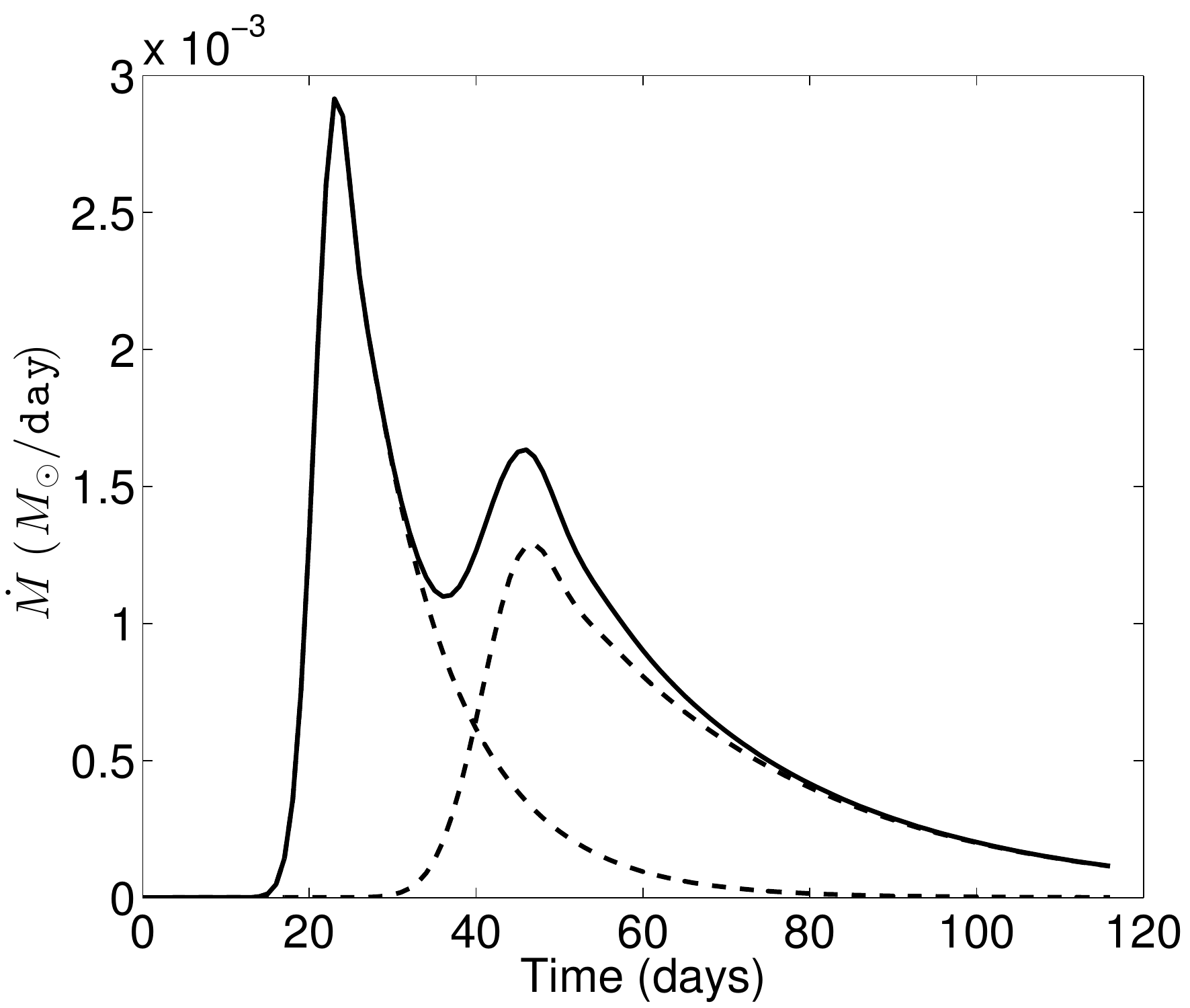}
  \caption{\label{fig:lightcurve} An illustration of the mass accretion rate for a typical double TDE described in the text, constructed using fitting formulae of \citet{GuillochonRamirezRuiz:2013}.} 
\end{figure}

As an illustrative example, we consider a binary with $0.40 \MSun$ and $0.27 \MSun$ components from our simulations.  The initial semimajor axis is $3.2 \RSun$ and the initial center-of-mass periapsis radius is $42 \RSun$.  The primary is disrupted first, but given the small binary separation, the secondary follows on nearly the same trajectory and is disrupted a few minutes later.  The secondary has a peak rise time of 46 days, compared to only 23 days for the primary.  We show an illustration of this double TDE in Fig.~\ref{fig:lightcurve}.  

When there is a single TDE, the surviving companion remains on a small angular momentum orbit.  Half of the surviving companions are on bound orbits, with a median return period of $\sim 50$ years.  The surviving companion may undergo a partial tidal disruption on the first periapsis passage, and be subsequently disrupted after returning to the MBH.

Binary collisions fall into two classes.  Some 30\% of mergers happen en route to the periapsis; the merger product arrives at periapsis after no more than two nominal dynamical timescales of the merged star, so we expect such merger products to be tidally disrupted even if the MBH periapsis is a few times the nominal tidal disruption radius of the merged star \citep[see also][]{Antonini:2011}.  The other 70\% of collisions happen after the stars pass periapsis, sometimes tens of tidal disruption radii away from the MBH; the close approach of binary components after periapsis passage had been pointed out by \citet{Sari:2010}.  These merger products have low angular momenta around the MBH, and most are likely to be tidally disrupted on future orbits as they undergo a random walk in angular momentum space.

\section{Discussion}\label{sec:conclusions}

We have carried out dynamics simulations of tidal interactions of stellar binaries with an MBH.  We found that both binary components are tidally disrupted in $18\%$ of our simulations.  Double TDE events may represent nearly $10\%$ of all stellar tidal disruptions.

Double TDEs may have a distinctive double-peak signature.  Intriguingly, \citet{Merloni:2015} recently interpreted the flare in AGN SDSS J015957.64+003310.5 as a possible TDE, but with an unexplained second peak in the light curve.  A prompt double TDE could provide a similar signal.  However, while Fig.~\ref{fig:lightcurve} can be viewed as an illustration of a double-peak light curve, we stress that we have not simulated the hydrodynamics of such an event.  Such simulations, including the possibility of orbital misalignments between tidal streams which we observe in some cases, are necessary to predict the full signature of double TDEs.  

A smaller fraction of our simulations resulted in single TDEs.  In half of these cases, the surviving star is left on a bound orbit with a period ranging from 6 months upwards, and a median period of half a century.  If the surviving star was partially tidally stripped during the initial encounter, it could lose more mass on a subsequent periapsis passage and produce a secondary flare.  The recent observation in IC 3599 of a repeated flare 20 years after a ROSAT event initially classified as a TDE \citep{Grupe:2015} could be representative of this scenario. 

Stellar collisions after periapsis passage can leave merger products on low angular momentum orbits around the MBH.  These merger products are likely to have strong magnetic fields \citep{Wickramasinghe:2014}.  Such magnetic fields could play a critical role in powering prompt jets \citep{GianniosMetzger:2011} such as observed in Swift J164449.3+573451 \citep{Bloom:2011,Burrows:2011,Levan:2011,Zauderer:2011}.  \citet{Tchekhovskoy:2014} argue that a magnetic flux $>10^{29}$ G cm$^2$ is required for prompt jet formation but is unlikely to be delivered by the star itself, and must be generated by the accretion-disc dynamo or be already present in the MBH vicinity. However, the magnetic field in the merger product may host $0.2$\% of the remnant's total energy \citep{Zhu:2015}, which implies a flux of $\sim 0.07 \sqrt{G} \Mstar=4 \times 10^{28} (\Mstar/M_\odot)$ G cm$^2$. This may serve as a useful seed for generating the field required for launching a jet.    

\acknowledgements
This research was supported by a Monash Research Acceleration grant (PI Y.~Levin).  IM is grateful for the hospitality of the Monash Center for Astrophysics.  Simulations in this paper made use of the collisional N-body code REBOUND which can be downloaded freely at \url{http://github.com/hannorein/rebound}.  We thank Fabio Antonini, Christopher Berry, James Guillochon and Clovis Hopman for helpful discussions and Hanno Rein for \texttt{REBOUND} advice.  


\bibliographystyle{hapj}


\end{document}